\documentclass[aps,pe,floatfix,superscriptaddress,twocolumn,showpacs]{revtex4}

\usepackage{epsfig}
\usepackage{amsmath}
\usepackage{graphicx,psfrag}
\usepackage{dcolumn}
\usepackage{bm}
\usepackage{slashed}

\newcommand{\beq}{\begin{equation}}
\newcommand{\eeq}{\end{equation}}
\newcommand{\bea}{\begin{eqnarray}}
\newcommand{\eea}{\end{eqnarray}}

\newcommand\fig[1]     {Fig.\,{\ref{#1}}}

\def\eq#1{(\ref{#1})}
\def\s0#1#2{\mbox{\small{$ \frac{#1}{#2} $}}}
\def\0#1#2{\frac{#1}{#2}}

\def\ord#1{{\cal O}(#1)}

\def\mr#1{{\mathrm{#1}}}

\begin{document}

\title{Magnetic particle hyperthermia: \\
Power losses under circularly polarized field in anisotropic nanoparticles} 

\author{I. N\'andori}
\affiliation{MTA-DE Particle Physics Research Group, H-4010 Debrecen P.O.Box 105, Hungary}
\affiliation{Institute of Nuclear Research, P.O.Box 51, H-4001 Debrecen, Hungary} 

\author{J. R\'acz}
\affiliation{Institute of Nuclear Research, P.O.Box 51, H-4001 Debrecen, Hungary}

\begin{abstract} 
The deterministic Landau-Lifshitz-Gilbert equation has been used to investigate 
the nonlinear dynamics of magnetization and the specific loss power in magnetic 
nanoparticles with uniaxial anisotropy driven by a rotating magnetic field, 
generalizing the results obtained for the isotropic case found in [P. F. de Ch\^atel, 
I. N\'andori, J. Hakl, S. M\'esz\'aros and K. Vad, J. Phys.: Condens. Matter {\bf 21}, 
124202 (2009)]. As opposed to many applications of magnetization reversal in 
single-domain ferromagnetic particles where losses must be minimized, in this 
paper, we study the mechanisms of dissipation used in cancer therapy by 
hyperthermia which requires the enhancement of energy losses. We show that 
for circularly polarized field, the loss energy per cycle is decreased by the 
anisotropy compared to the isotropic case when only dynamical effects are 
taken into account. Thus, in this case, in the low frequency limit, a better heating 
efficiency can be achieved for isotropic nanoparticles. The possible role of thermal 
fluctuations is also discussed. Results obtained are compared to experimental data. 
\end{abstract}

\maketitle

\section{Introduction}
\label{sec_intro}
The nonlinear dynamics of the magnetization in single-domain ferromagnetic 
nanoparticle systems has been the subject of an intense study and it is at date 
a challenging issue. Examples are ferromagnetic resonance, switching of 
magnetization, data storage based on magnetic devices, spintronics etc. The 
applications which are strongly related to the present work are the following: 
ferrofluids, magnetic resonance imaging (MRI)  and other biomedical applications, see 
e.g. \cite{ferrofluid,ferrohydro,Petrova,biomedical,Raikher2011,Cantillon, Ahsen2010}. 
While in most cases the loss energy per cycle has to be minimized, in cancer 
therapy by hyperthermia the goal is to enhance the heating efficiency of magnetic 
nanoparticles driven by an external magnetic field, preferably 
inside the malignant tumours. The common practice is to use a linearly polarized 
external magnetic field alternating at a frequency of the order of $10^5$ Hz. 
Indeed, in case of the linearly polarized applied field, the optimization of loss 
energy with respect to the amplitude and frequency of the external field has been 
studied in detail \cite{biomedical,CoFa2002,Poperechny2010}. It is natural to ask 
what is the dependence of the specific absorption rate on the nature  of polarization, 
i.e. whether a better heating efficiency can be achieved by a circularly polarized 
applied field \cite{Chatel2009,Raikher2011,Cantillon,Ahsen2010}. The study of 
dynamical effects of circularly polarized field has received a considerable attention
\cite{Bertotti2001,SunWang,Denisov2006,Denisov2006prl,Denisov_thermal}.
Power losses for isotropic nanoparticles under rotating field have also been 
investigated in the presence (see e.g. \cite{Cantillon,Raikher2011}) or in the
absence (see e.g. \cite{Chatel2009}) of thermal effects, but no systematic analysis
have been performed in order to investigate the effect of anisotropy on the energy
absorption of nanoparticles in the low frequency limit suitable for hyperthermia.

The relaxation and the loss energy of a single isotropic magnetic nanoparticle 
has been considered under circularly polarized applied field in \cite{Chatel2009} 
when no thermal effects were included. In the low frequency limit, the loss energy 
per cycle was found to be larger in case of the linearly polarized applied field as 
compared to the circularly polarized one. Thermal effects for isotropic system were 
studied in detail in \cite{Raikher2011}. In the limit of low frequency, the linearly 
polarized field was found to produce more heat power for higher temperatures, too. 
However, recent experimental results \cite{Ahsen2010} show a different picture; 
the linearly and the circularly polarized external field produced an equal heat 
power at least for low frequencies. Since, the immobility and the aggregation of 
particles into chains is a known feature of ferrofluids when the sample becomes 
very anisotropic, it is a natural question to ask whether the anisotropy can be 
responsible for the discrepancy between the theory and the experiment.

The goal of this paper is twofold. On the one hand, we consider the role of 
anisotropy in the possible enhancement of heat power of magnetic nanoparticles 
(in the absence of thermal effects) by generalizing the results obtained for the 
isotropic case found in \cite{Chatel2009}. On the other hand, we study whether 
the anisotropy can be used to explain the experimental results of \cite{Ahsen2010}.  
The possible role of thermal fluctuations is also discussed.

The paper is organized as follows. In Sec~\ref{sec_neel}, the deterministic
Landau-Lifshitz-Gilbert equation has been given in case of uniaxial anisotropy 
suitable for the description of magnetization dynamics for single magnetic 
nanoparticles (in the temperature range far from the Curie temperature). The 
specific loss power and loss energy is studied in case of the circularly polarized 
applied field in Sec~\ref{sec_circ}. Known results of the isotropic case is briefly 
summarized and compared to the findings of the present work done for nanoparticles 
with uniaxial anisotropy. In Sec~\ref{sec_lin}, we study the linearly polarized applied 
field in the limit of large anisotropy and in Sec~\ref{sec_thermal} the possible 
modification of the findings by thermal fluctuations is discussed. Finally, 
Sec~\ref{sec_sum} stands for the summary.

\section{Landau-Lifshitz-Gilbert equation}
\label{sec_neel}

In order to study the energy losses under repeated magnetization reversal, one 
can distinguish two different processes related to the mobility of magnetic 
particles in ferrofluids. Either the magnetic moment rotates within the particle 
(N\'eel regime \cite{Neel}) or the particle rotates as a whole (Brown regime). 
Another way to classify various types of relaxation mechanisms is related to 
the temperature. For example, if one considers relaxation far from the Curie 
temperature then the magnetization process in a single-domain particle can 
be well described by means of the Landau--Lifshitz \cite{LaLi1935} equation 
which is mathematically equivalent to Gilbert's one \cite{Gilbert} with the 
appropriate definition of its coefficient \cite{Br1979,Gilbert_llg}. This is referred 
as the Landau--Lifshitz--Gilbert (LLG) equation. For a complete description of 
relaxations close the Curie temperature, thermal effects should be incorporated. 
However, in some cases (such as a rotating applied field) the LLG equation 
provides us reliable results on energy losses even at higher temperatures. 
 
In this work the focus is on the relaxation based on dynamical effect obtained 
by the LLG equation for anisotropic nanoparticles. 
We argue that at least for small and relatively large anisotropy, findings of the 
present work can be used to study energy losses in a temperature range relevant 
for hyperthermia. An important feature of the LLG equation is that  the magnetization 
vector's magnitude does not change under the influence of the external field. Thus, 
it is convenient to rewrite it in terms of the unit vector ${\bf M} = {\bf m}/m_S$ where 
$m_S$ is the saturation magnetization. Then the LLG equation reads as
\begin{equation}
\label{LLG}
\frac{\mr{d}}{\mr{d}t} {\bf M} = -\gamma' [{\bf M \times H_{\mr{eff}}}] 
+ \alpha' [[{\bf M\times H_{\mr{eff}}]\times M}].
\end{equation}
with the coefficients $\gamma'  = \mu_0 \gamma_0/(1+\alpha^2)$ and 
$\alpha' = \mu_0^2 \gamma_0^2 \eta m_S/(1+\alpha^2)$ where $\mu_0$ is 
the permeability of free space, $\gamma_0$ is the gyromagnetic ratio and 
$\eta$ is the dimensionfull and $\alpha = \mu_0 \gamma_0 \eta m_S$ the
dimensionless damping constant. Let us  introduce an effective gyromagnetic 
ratio $\gamma = \gamma_0/(1+\alpha^2)$. Then the parameters of the LLG
equation \eq{LLG} can be rewritten as $\gamma' = \mu_0 \gamma$ and 
$\alpha' = \mu_0 \gamma \alpha$.  It is important to note that the effective 
gyromagnetic ratio $\gamma$ used in this paper is positive as opposed to 
the negative parameter of \cite{Chatel2009}. The cross denotes the vector 
product and the effective magnetic field acting on the magnetization ${\bf M}$ 
is defined as
\begin{equation}
{\bf H}_{\mr{eff}} = {\bf H}_{\mr{ext}} + {\bf H}_{\mr{aniso}}
\end{equation}
with the alternating or circulating applied (external) field 
${\bf H}_{\mr{ext}}$ and the anisotropy field ${\bf H}_{\mr{aniso}}$.
Let us note, that in this work all considerations have been done in the 
absence of a static field.

\section{Circularly polarized applied field}
\label{sec_circ}

In this section we discuss the solution of the LLG equations \eq{LLG} obtained 
for an immobile single-domain (isotropic and anisotropic) magnetic particle 
under circularly polarized, i.e. rotating applied (external) magnetic field. The 
applied field is assumed to rotate in the $xy$-plane with an angular frequency 
$\omega$
\begin{eqnarray}
\label{rot_field}
{\bf H_{\mr{ext}}} &=& 
\frac{\omega_L}{\gamma'} \, \, (\cos(\omega t), \sin(\omega t), 0),
\end{eqnarray}
where $\omega_L = \gamma' \vert{\bf H}\vert$ is the Larmor frequency.
(The angular velocity vector is perpendicular the the $xy$ plane.) For the
sake of simplicity we consider particles with uniaxial anisotropy where
the easy axis of the magnetization is chosen to be the z-axis, i.e. the 
anisotropy field is defined as
\begin{eqnarray}
\label{aniso_rot_field}
{\bf H_{\mr{aniso}}} &=& 
\frac{\omega_L}{\gamma'} \, \, (0, 0, \lambda_{\mr{eff}} M_z),
\end{eqnarray}
where $M_z$ is the z-component of the magnetization vector and the
parameter $\lambda_{\mr{eff}}$ describes the strength of the anisotropy.
With this particular choice of the anisotropy field the arrangement used in the 
paper is identical to that investigated in \cite{Bertotti2001,Denisov2006}. 
For a more detailed analysis, specially when the effect of thermal fluctuations
is also taken into account, the easy axis of magnetization has to be chosen 
arbitrarily (for linearly polarized case see e.g. \cite{stochastic_llg_lin}) 
but this is out of the scope of the present work.

It is convenient to use a coordinate system in which the steady state solution
of the LLG equation \eq{LLG} has the simplest form: a time-independent 
magnetization vector \cite{Chatel2009,Bertotti2001,Denisov2006}. The 
transformation is done by an appropriate rotation \cite{Chatel2009}
\begin{align} 
\label{rot1}
\underline{\underline{\mr{O}_1}} = 
\left(\begin{array}{ccc} 
+\cos(\omega t) &\hspace*{0.2cm} +\sin(\omega t) &\hspace*{0.2cm} 0 
\\
-\sin(\omega t) &\hspace*{0.2cm} +\cos(\omega t) &\hspace*{0.2cm} 0
\\
0               &\hspace*{0.2cm} 0              &\hspace*{0.2cm} 1 
\end{array} \right)
\end{align}
which transforms the LLG equation into a coordinate system, which rotates 
around the $z$ axis with the applied magnetic field. The transformed $z$ 
axis points then in the direction of the angular velocity vector ${\bf \omega}$. 
Denoting the Cartesian coordinates of the transformed magnetization
\begin{eqnarray}
(u_x ,u_y ,u_z ) = \underline{\underline{\mr{O}_1}}{\bf M},
\end{eqnarray}
the LLG equation \eq{LLG} can be written as
\begin{eqnarray}
\label{rot_LLG_descartes}
\frac{\mr{d} u_x}{\mr{d}t} &=& 
\omega  u_y + \alpha_N u_y^2 + \alpha_N u_z^2   \nonumber \\
&&-\omega_L \lambda_{\mr{eff}} u_y u_z
- \alpha_N \lambda_{\mr{eff}} u_x u_z^2,
\nonumber \\
\frac{\mr{d} u_y}{\mr{d}t} &=& 
-\omega u_x - \omega_L u_z -\alpha_N u_x u_y \nonumber \\
&&+\omega_L \lambda_{\mr{eff}} u_x u_z 
- \alpha_N \lambda_{\mr{eff}} u_y u_z^2,
\nonumber \\
\frac{\mr{d} u_z}{\mr{d}t} &=& \omega_L u_y - \alpha_N u_x u_z 
+ \alpha_N  \lambda_{\mr{eff}} (1-u_z^2 ) u_z, 
\end{eqnarray}
where $\alpha_N = \alpha'  \vert{\bf H}\vert$ is introduced. Let us note
that $\lambda_{\mr{eff}}$ is dimensionless but $\alpha_N$, $\omega$
and $\omega_L$ are of dimension $1/s$. However, by introducing a 
dimensionless time $\tilde t = t/t_0$ all the frequency parameters can 
be rewritten as dimensionless quantities such as 
$\tilde \alpha_N = \alpha_N t_0$ etc. For the sake of simplicity we keep 
the original notation (without the tilde superscript) but the time and 
consequently the frequencies are considered as dimensionless 
parameters. Since the LLG equation retains the magnitude of the 
magnetization vector, only two of the Cartesian components are 
independent (${\bf u}$ is a unit vector in the rotating frame). In order to 
describe the orientation of the magnetization let us introduce angles 
following the definition of \cite{Bertotti2001},
\begin{eqnarray}
\label{def_angles}
u_x &=& \sin\theta \cos\phi, \nonumber \\
u_y &=& -\sin\theta \sin\phi, \nonumber \\
u_z &=& \cos\theta.
\end{eqnarray}
Thus in the rotating frame the LLG equation obtained for the three 
Cartesian coordinates \eq{rot_LLG_descartes} reduces to a set of
differential equations for the two angles 
\begin{eqnarray}
\label{rot_LLG_angles}
\frac{\mr{d} \theta}{\mr{d}t} &=& 
\omega_L \sin\phi + \alpha_N \cos\theta \cos\phi 
- \alpha_N \lambda_{\mr{eff}} \sin\theta \cos\theta,
\nonumber \\
\frac{\mr{d} \phi}{\mr{d}t} &=& 
\omega_L \cos\phi \frac{\cos\theta}{\sin\theta} + w 
- \alpha_N \frac{\sin\phi}{\sin\theta} 
- \omega_L \lambda_{\mr{eff}} \cos\theta.
\end{eqnarray}
Let us note that the differential equations for $\theta$ and $\phi$ in 
\eq{rot_LLG_angles} are identical to Eqs.(2,3) of \cite{Bertotti2001} 
in case of vanishing static field ($h_{az} =0$) if one makes the following 
identifications $\omega_L \equiv h_{a\perp}/(1+\alpha^2)$, 
$\alpha_N \equiv \alpha h_{a\perp}/(1+\alpha^2)$ and
$\lambda_{\mr{eff}} \equiv  \kappa_{\mr{eff}}/h_{a\perp}$.

\subsection{Isotropic case}
\begin{figure}[ht] 
\begin{center} 
\epsfig{file=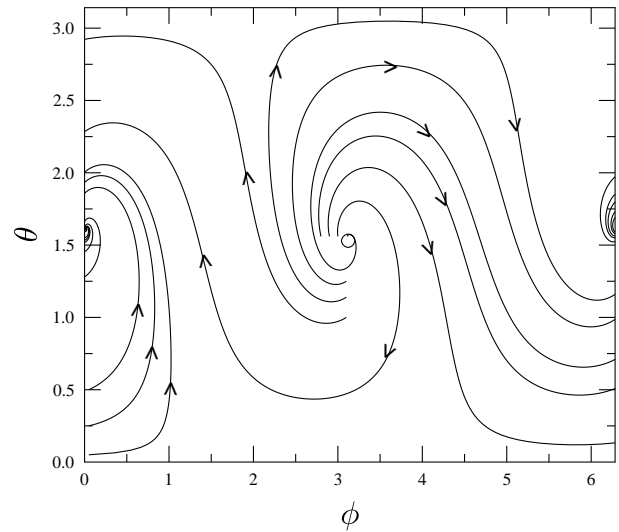,width=8.0 cm}
\caption{
\label{fig1}
Phase portrait in the rotating frame obtained by solving the LLG equation 
\eq{rot_LLG_angles} for the isotropic case ($ \lambda_{\mr{eff}} = 0$) with
the parameters $\alpha_N = 0.1, \omega = 0.01, \omega_L = 0.2$. The 
circle indicates the repulsive fixed point at $\phi = 3.12$, $\theta = 1.53$.
The attractive fixed point is at $\phi = 0.02$, $\theta = 1.61$.} 
\end{center}
\end{figure}

In this subsection we briefly summarize the results obtained for the isotropic 
case in \cite{Chatel2009}. Numerical solutions of Eq.~\eq{rot_LLG_angles}
derived for the isotropic case ($ \lambda_{\mr{eff}} = 0$) with various initial 
conditions are plotted in \fig{fig1} (for a set of parameters given in the figure 
caption). 
%
%
Two fixed points, a repulsive (circle) one and an attractive one appear in 
the phase diagram. Both can be determined by the analytical solution of 
the algebric fixed point equation derived from \eq{rot_LLG_descartes} in 
case of vanishing anisotropy i.e. for $ \lambda_{\mr{eff}} = 0$. The solution 
for the attractive fixed point is (see Eq. (25) in \cite{Chatel2009}),
\begin{eqnarray}
\label{ux0_uy0}
&u_{x0} = \sqrt{\frac{\alpha_N^2 -\omega_L^2 -\omega^2 + 
\sqrt{4\alpha_N^2 \omega_L^2 
+ (\alpha_N^2 -\omega_L^2 -\omega^2)^2}}{2\alpha_N^2}},
\nonumber \\
&u_{y0} = - \frac{\alpha_N^2 +\omega_L^2 +\omega^2 - 
\sqrt{4\alpha_N^2 \omega_L^2 
+ (\alpha_N^2 -\omega_L^2 -\omega^2)^2}}{2\omega\alpha_N},
\nonumber \\
&u_{z0} = -\sqrt{1- u_{x0}^2 -u_{y0}^2},
\end{eqnarray}
(in case of the repulsive fixed point $u_{x0}$ and $u_{z0}$ have been
multiplied by -1). The attractive fixed point of the LLG equation in the 
rotating frame corresponds to the stable steady state solution obtained in 
the laboratory frame (Eq. (24) in \cite{Chatel2009})
\begin{eqnarray}
\label{rot_sol_relax_spec}
M_{x}(t) &=& 
u_{x0} \cos(\omega t) - u_{y0} \sin(\omega t),
\nonumber \\
M_{y}(t) &=&
 u_{x0} \sin(\omega t) + u_{y0} \cos(\omega t),
\nonumber \\
M_{z}(t) &=& u_{z0}.
\end{eqnarray}
The steady state solution enables us to calculate the energy loss for a single 
particle. The energy dissipated in a single cycle can be calculated as 
(based on Eq. \eq{rot_sol_relax_spec})
\begin{eqnarray}
\label{def_loss}
E = \mu_0 m_S \int_{0}^{\frac{2\pi}{\omega}} \mr{d}t 
\left({\bf H} \cdot \frac{d{\bf M}}{dt} \right)
= \mu_0 2\pi m_S H (-u_{y0}),
\end{eqnarray}
(see also Eq.~(32) in \cite{Chatel2009}) which has the form in the low-frequency 
limit, $\omega \ll \alpha_N$,
\begin{eqnarray}
\label{loss_iso}
E =  2 \pi \mu_0 m_S H  
\left[\frac{\alpha_N \omega}{\omega_L^2 + \alpha_N^2}
-\frac{\alpha_N  \omega_L^2 \omega^3}{(\omega_L^2 + \alpha_N^2)^3}
+\ord{\omega^5}  \right].
\end{eqnarray}
Let us note that in Ref.~\cite{Chatel2009} the energy loss per cycle of isotropic 
nanoparticles obtained by oscillating and rotating external fields have been
analyzed in the absence of thermal effects, see e.g. Fig. 2 in \cite{Chatel2009}. 
It was shown that in the low-frequency limit the energy loss per cycle was found 
to be larger in the linearly polarized case. In order to consider the role of thermal 
fluctuations in case of isotropic samples let us compare the findings of 
\cite{Raikher2011} (where thermal effects were included) to \cite{Chatel2009}. 
Dashed lines on Fig. 6 of \cite{Raikher2011} correspond to the limit $T \to 0$ and 
agree to the findings plotted in Fig. 2a of \cite{Chatel2009} qualitatively. The 
important result is that in the limit of low frequency, the linearly polarized field 
was found to produce more heat power both for $T=0$ and for $T \neq 0$.

\subsection{Anisotropic case}

In order to study the role of anisotropy let us follow the strategy applied
in the isotropic case. Namely, we calculate the loss energy per cycle by 
determining the attractive fixed point solution of \eq{rot_LLG_descartes}
for non-vanishing anisotropy ($ \lambda_{\mr{eff}} \neq 0$) which reads
\begin{align}
\label{fp1}
&0 = \omega  u_y + \alpha_N u_y^2 + \alpha_N u_z^2
-\omega_L \lambda_{\mr{eff}} u_y u_z
- \alpha_N \lambda_{\mr{eff}} u_x u_z^2,
\nonumber \\
&0 = 
-\omega u_x - \omega_L u_z -\alpha_N u_x u_y 
+\omega_L \lambda_{\mr{eff}} u_x u_z 
- \alpha_N \lambda_{\mr{eff}} u_y u_z^2,
\nonumber \\
&0 = \omega_L u_y - \alpha_N u_x u_z 
+ \alpha_N  \lambda_{\mr{eff}} (1-u_z^2 ) u_z,
\end{align}
which can be reduced for the following equation for $u_{z}$,
\begin{align}
\label{fp4}
0 = \frac{1-u_z^2}{u_z^2} \left( 
\left[u_z - \frac{r \, \omega_L}{\lambda_{\mr{eff}}}\right]^2 
+ \left[u_z \frac{r \, \alpha_N}{\lambda_{\mr{eff}}}\right]^2
\right) - \frac{1}{\lambda_{\mr{eff}}^2} 
\end{align}
with $r = \omega/(\omega_L^2 + \alpha_N^2)$. Let us note that 
Eq. \eq{fp4} is identical to Eq. (3.8) of \cite{Denisov2006} (with 
zero static field $\tilde H =0$ and with $\rho = 1$) if the following 
identifications are used: ${\tilde h} \equiv 1/\lambda_{\mr{eff}}$,
$\kappa \equiv  r \omega_L /\lambda_{\mr{eff}}$ and 
$\lambda \equiv \alpha_N/\omega_L$. The switching of the nanoparticle 
magnetic moments and the dynamical effects under the action of rotating 
field have been studied in \cite{Denisov2006} with great details and further 
considered in \cite{Denisov2006prl,Denisov_thermal}, however energy 
losses were not calculated which is the goal of the present work.

\begin{figure}[ht] 
\begin{center} 
\epsfig{file=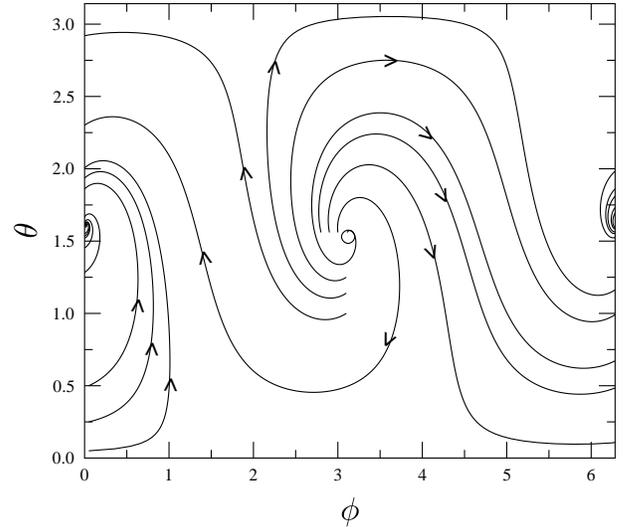,width=8.0 cm}
\caption{
\label{fig2}
Phase portrait in the rotating frame obtained by solving the LLG equation
\eq{rot_LLG_angles} in the limit of low-frequency and small anisotropy
for the parameters $\alpha_N = 0.1, \omega = 0.01, \omega_L = 0.2$ and
$\lambda_{\mr{eff}} = 0.05$. It is similar but not identical to \fig{fig1}.} 
\end{center}
\end{figure}

\begin{figure}[ht] 
\begin{center} 
\epsfig{file=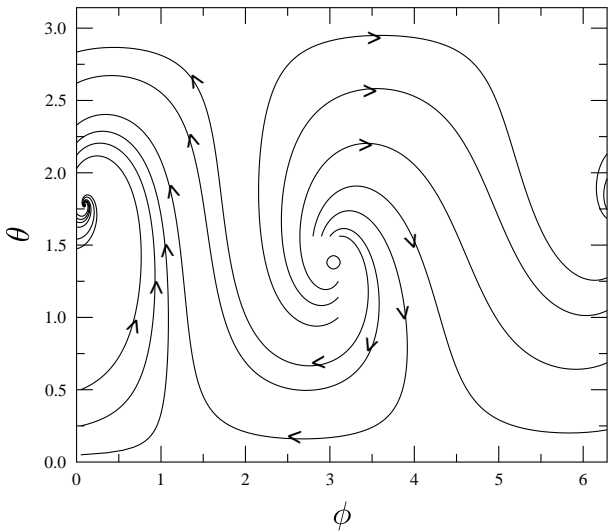,width=8.0 cm}
\caption{
\label{fig3}
Phase portrait in the rotating frame obtained by solving the LLG equation
\eq{rot_LLG_angles} (medium-frequency and small anisotropy) for the 
parameters $\alpha_N = 0.1, \omega = 0.05, \omega_L = 0.2$ and
$\lambda_{\mr{eff}} = 0.05$. According to numerical results, the attractive 
fixed point is at $\phi = 0.097995$, $\theta = 1.77908$ which gives 
$u_{y0} = -0.0957$.}
\end{center}
\end{figure}

Eq.~\eq{fp4} can be solved analytically which is used to calculate all the 
Cartesian coordinates of the attractive fixed point (or fixed points) in the 
rotating frame. If $u_{y0}$ (y-axis component of the attractive fixed point) 
is known the energy loss per cycle can be evaluated similarly to the 
isotropic case. For biomedical applications such as hyperthermia the 
the low-frequency limit is relevant. Therefore, let us first consider the 
low-frequency $\omega \ll \alpha_N$ and small anisotropy 
$\lambda_{\mr{eff}}\ll 1$ limit where one finds a single attractive fixed
point with
\begin{align}
\label{fp_low-freq_small-aniso}
u_{y0} \approx  -
\frac{\alpha_N \omega}{\omega_L^2 + \alpha_N^2}
+ \frac{\alpha_N  \omega_L^2 \omega^3}{(\omega_L^2 + \alpha_N^2)^3}
(1+2 \lambda_{\mr{eff}}).
\end{align}
Inserting \eq{fp_low-freq_small-aniso} into the expression of the energy
loss per cycle \eq{def_loss} one finds,
\begin{align}
\label{loss_low-freq_small-aniso}
E \approx 2 \pi \mu_0 m_S H \left[
\frac{\alpha_N \omega}{\omega_L^2 + \alpha_N^2}
- \frac{\alpha_N  \omega_L^2 \omega^3}{(\omega_L^2 + \alpha_N^2)^3}
(1+2 \lambda_{\mr{eff}}) \right],
\end{align}
which shows that in the low-frequency limit, the small anisotropy does not 
modify the energy loss per cycle obtained for the isotropic case. At higher 
frequencies in case of small anisotropy a decrease is observed in the energy 
dissipated in a single cycle compared to the isotropic case. This analytic 
result is supported by the numerical integration of Eq.~\eq{rot_LLG_angles},
(low-frequency, small anisotropy), see \fig{fig2}, which is very similar to 
the isotropic case, see \fig{fig1}. 
%
%
Figures \fig{fig1} and \fig{fig2} are similar but not identical. Trajectories
of the two figures, far from the fixed points differ from each other (but the 
deviation is small). However, for low-frequency and small anisotropy, the 
attractive and the repulsive fixed points of \fig{fig1} and \fig{fig2} coincide. 
The stability analysis done in \cite{Denisov2006} also confirms the existence 
of a single attractive fixed point in this regime of the parameter space. 
The difference between the positions of the attractive fixed point obtained
for the isotropic and anisotropic cases is more recognizable at higher
angular frequencies. For example, according to the approximate expression 
\eq{fp_low-freq_small-aniso}, the attractive fixed point of \fig{fig3} is at 
$u_{y0} = -0.0956$, numerical results give $u_{y0} = -0.0957$ 
and the corresponding isotropic case (see, Eq. \eq{ux0_uy0}) gives 
$u_{y0} = -0.0961$. 
%
%
Thus the energy loss per cycle (which is related to $-u_{y0}$) is decreased 
for the set of parameters used in \fig{fig3} compared to the isotropic case 
(for the same $\omega$, $\omega_L$ and $\alpha_N$). 

\begin{figure}[ht] 
\begin{center} 
\epsfig{file=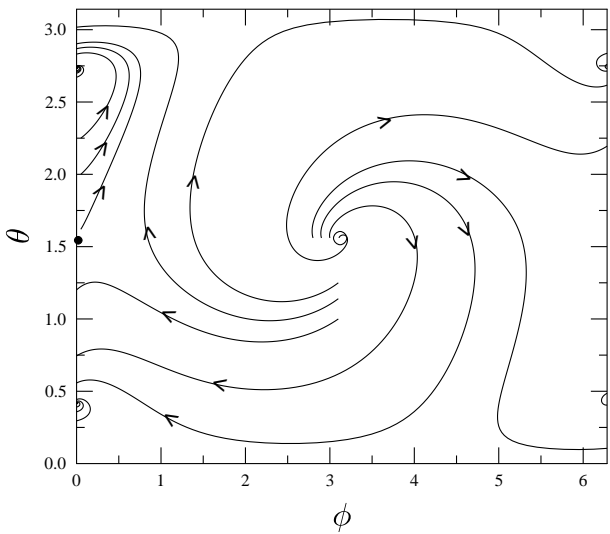,width=8.0 cm}
\caption{
\label{fig4}
Phase portrait in the rotating frame obtained by solving the LLG equation
\eq{rot_LLG_angles} in the limit of low-frequency and large anisotropy for 
the parameters $\alpha_N = 0.1, \omega = 0.01, \omega_L = 0.2$ and
$\lambda_{\mr{eff}} = 2.5$. Two attractive fixed points appear in the figure.
The repulsive fixed point and the saddle point are indicated by a circle
and a solid circle, respectively.} 
\end{center}
\end{figure}

\begin{figure}[ht] 
\begin{center} 
\epsfig{file=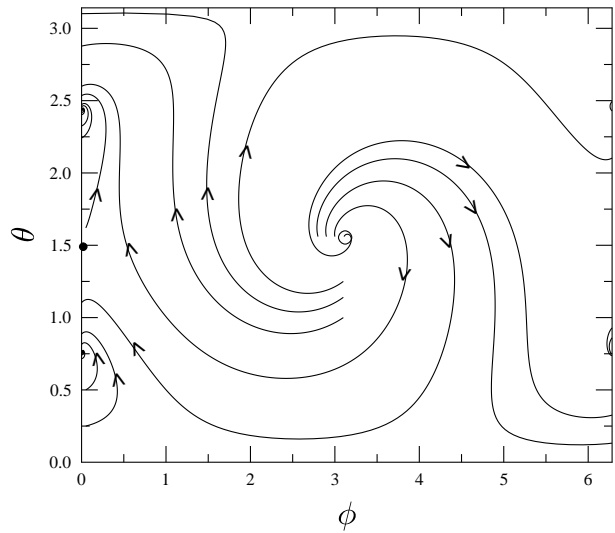,width=8.0 cm}
\caption{
\label{fig5}
Phase portrait in the rotating frame obtained by solving the LLG equation
\eq{rot_LLG_angles} (low-frequency, large anisotropy) for the parameters 
$\alpha_N = 0.1, \omega = 0.01, \omega_L = 0.2$ and 
$\lambda_{\mr{eff}} = 1.5$. The attractive fixed points are closer to the 
equator as compared to \fig{fig4} where the anisotropy was larger.} 
\end{center}
\end{figure}

\begin{figure}[ht] 
\begin{center} 
\epsfig{file=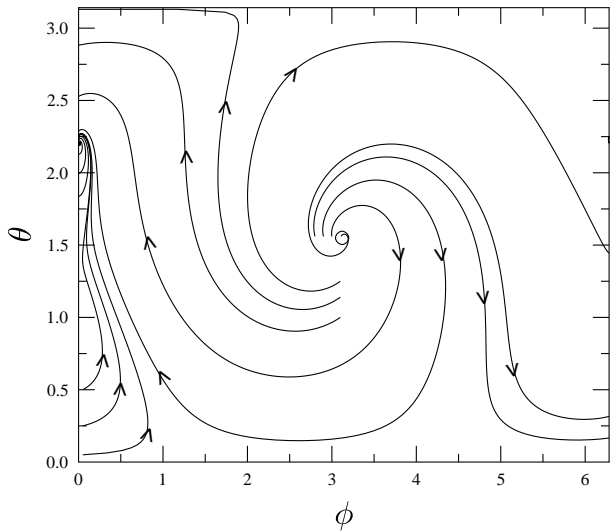,width=8.0 cm}
\caption{
\label{fig6}
Phase portrait in the rotating frame obtained by solving the LLG equation
\eq{rot_LLG_angles} in the limit of low-frequency, slightly below the critical
value of anisotropy. The parameters  are $\alpha_N = 0.1, \omega = 0.01, 
\omega_L = 0.2$ and $\lambda_{\mr{eff}} = 1.175$. There is only a single 
attractive fixed point below the equator (large $\theta$). The other attractive
fixed point (above the equator, small $\theta$) of the large anisotropic case 
and also the saddle point vanish in this regime of the parameter space.}
\end{center}
\end{figure}

Let us consider the low-frequency but large-anisotropy limit. In this case
one finds two attractive fixed points, see \fig{fig4}. 
%
%
The attractive fixed points are situated above (up) and below (down) the 
equator. In the strong anisotropy limit their $\phi$-components are the 
same and their $\theta$-components are symmetric to the equator, thus their 
Cartesian components are $u_{x0}^{\mr{up}} = u_{x0}^{\mr{down}}$,
$\vert u_{z0}^{\mr{up}}\vert = \vert u_{z0}^{\mr{down}} \vert$ and 
$u_{y0}^{\mr{up}} = u_{y0}^{\mr{down}}$. Therefore, both attractive fixed 
points have the same y-components in the rotating frame which reads as
\begin{align}
\label{fp_low-freq_large-aniso}
u_{y0} \approx  - \frac{\alpha_N \omega}{\omega_L^2 + \alpha_N^2} 
\frac{1}{\lambda_{\mr{eff}}^2}.
\end{align}
Inserting \eq{fp_low-freq_large-aniso} into the expression of energy
loss per cycle \eq{def_loss} one finds,
\begin{align}
\label{loss_low-freq_large-aniso}
E \approx 2 \pi \mu_0 m_S H \left[
\frac{\alpha_N \omega}{\omega_L^2 + \alpha_N^2} 
\frac{1}{\lambda_{\mr{eff}}^2}
\right],
\end{align}
which vanishes for $\lambda_{\mr{eff}} \to\infty$. In the limit of extreme 
large anisotropy there is no room for energy dissipation since the
magnetization is aligned to the easy axis independently of the applied
field. Indeed, for $\lambda_{\mr{eff}} \to\infty$ the xy-plane components
tend to zero $u_{x0} \to 0$, $u_{y0} \to0$ and $u_z \to \pm 1$. In the 
($\phi$-$\theta$) plane the attractive fixed points should tend to the "poles", 
i,e, $\phi = 0$ and $\theta = 0,\pi$. If the anisotropy is decreased, they 
"move away" from the poles and tend to the equator, see \fig{fig5}.
%
%
A critical value for the anisotropy parameter can be identified where one of 
the attractive fixed point (the one which corresponds to small $\theta$, i.e.
which lies above the equator \eq{def_angles}) vanishes. The other
attractive fixed point remains always below the equator. The phase
diagram obtained in the low-frequency limit, slightly below the critical
value of anisotropy is plotted in \fig{fig6}. 
%
%

\begin{figure}[ht] 
\begin{center} 
\epsfig{file=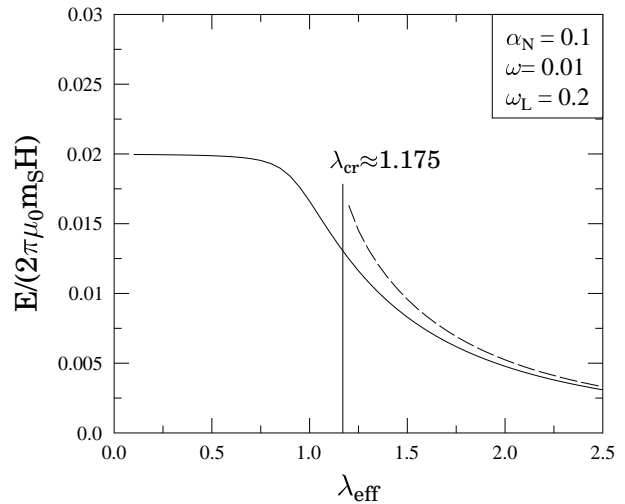,width=8.0 cm}
\caption{
\label{fig7}
The energy loss is plottes versus the anisotropy $\lambda_{\mr{eff}}$. 
There are two scaling regimes, the one is at small, the other is at large
anisotropy which are separated by the critical value $\lambda_{\mr{cr}}$. 
The full line corresponds to energy loss obtained at the stable fixed 
point below the equator and the dashed line represents energy loss 
at the fixed point above the equator which is stable only for 
$\lambda_{\mr{eff}} > \lambda_{\mr{cr}}$.} 
\end{center}
\end{figure}

Let us consider the loss energy as a function of the anisotropy parameter.
In \fig{fig7} the loss energy is plotted versus $\lambda_{\mr{eff}}$ (for a set
of parameters given in the figure) and it shows that in case of a rotating field 
the loss energy is obtained to be a monotonic function of the anisotropy 
parameter.
%
%
Thus, according to our results, the anisotropy (where the easy axis is 
perpendicular to the rotating external field) cannot be used to increase 
the heating efficiency of magnetic nanoparticles in the low-frequency 
limit.

\begin{figure}[ht] 
\begin{center} 
\epsfig{file=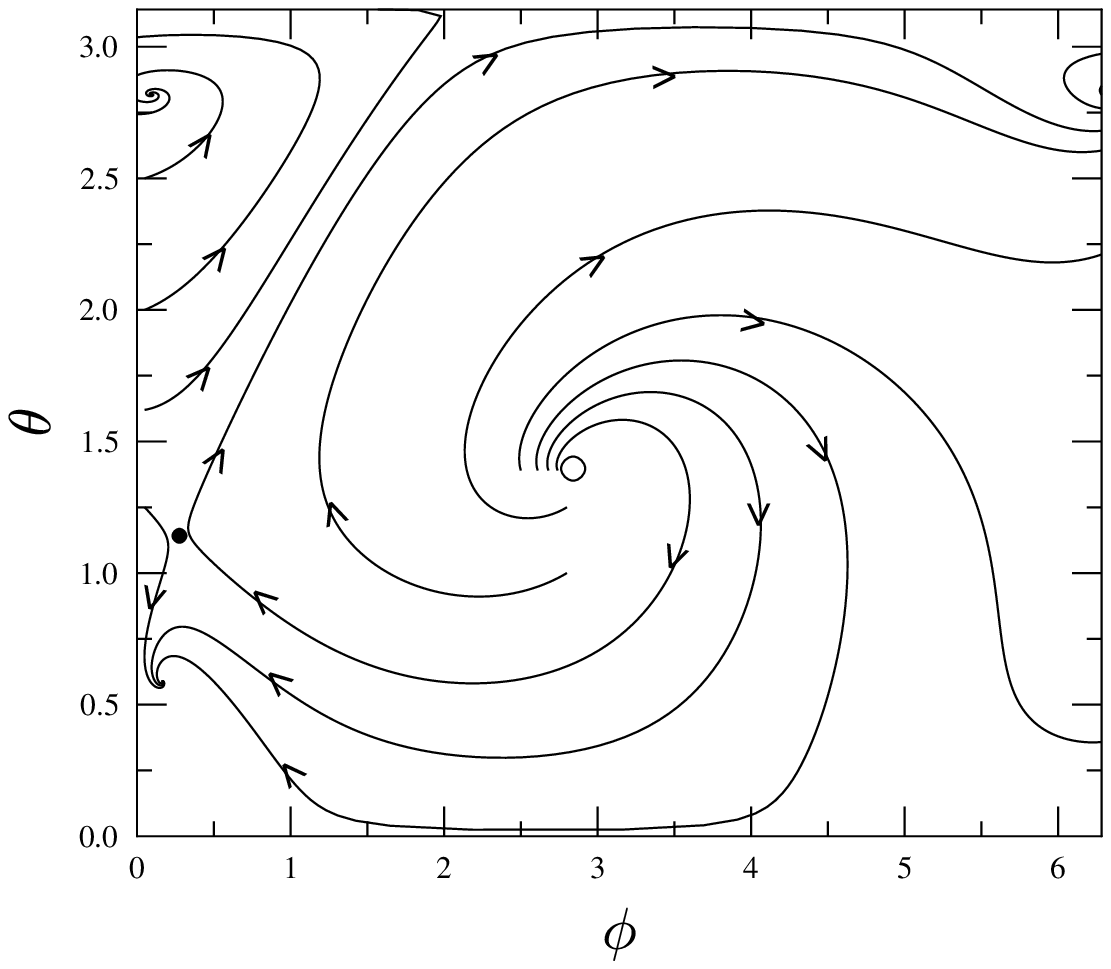,width=8.0 cm}
\caption{
\label{fig8}
Phase portrait in the rotating frame obtained by solving the LLG equation
\eq{rot_LLG_angles} in the limit of high-frequency and large anisotropy 
for the parameters $\alpha_N = 0.1, \omega = 0.15, \omega_L = 0.2$ and 
$\lambda_{\mr{eff}} = 2.5$. There are two attractive fixed points in the figure
similarly to the low-frequency case, \fig{fig4}. Here the attractive fixed points 
have different y-components, thus the corresponding energy losses also 
differ from each other.} 
\end{center}
\end{figure}

For the sake of completeness let us consider the high-frequency limit,
although it is out of the scope of the present work (irrelevant in case of 
cancer therapy by hyperthermia), hence, we do not study this in detail. 
The phase diagram obtained in the high-frequency and large anisotropy 
limit is shown in \fig{fig8} and can be compared to the one obtained for
low frequencies (with the same anisotropy), see \fig{fig4}.
%
%
The similarity between the high and low frequency cases is that two
attractive fixed points appear. However, in the high-frequency case
they have different coordinates in the ($\phi$-$\theta$) plane. Thus, the 
energy losses correspond to the attractive fixed points differ from each other.

In summary we conclude that the 
uniaxial anisotropy (where the easy axis is perpendicular to the rotating 
external field) either does not modify the energy loss per cycle (in case of 
small anisotropy) or the energy dissipated is decreased as compared to the 
isotropic case.

\section{Linearly polarized applied field in the limit of large anisotropy}
\label{sec_lin}

One of the goal of this paper is to investigate the role of anisotropy in 
the possible enhancement of heat efficiency of magnetic nanoparticles
driven by a rotating magnetic field. On the one hand, in the previous 
section it was obtained that in case of circularly polarized 
applied field if the uniaxial anisotropy (perpendicular to the rotating 
external field) has been taken into account, the energy loss per cycle 
either remains unchanged (small anisotropy) or decreased (large 
anisotropy). On the other hand, in Ref.~\cite{Chatel2009} it was 
shown that for isotropic nanoparticles the linearly polarized external 
field provides us a larger heat power in the limit of low frequency. 
Furthermore, the latter statement was shown \cite{Raikher2011} to be 
reliable in the presence of thermal effect, too. Thus, the circularly polarized 
applied field cannot be used to achieve a better heating efficiency by 
nanoparticle systems (neither for isotropic, nor for anisotropic nanoparticles 
with uniaxial anisotropy) if the effect of thermal fluctuations is negligible. The 
role of thermal effects is discussed in \ref{sec_thermal} where it is argued 
that the possible modification of the above finding by thermal fluctuations can 
only be expected in case of moderate anisotropy. Thus, the results of the
present work indicates that the heating efficiency cannot be increased by the 
rotating field at least for small and very large anisotropy in the limit of low
frequencies (independently whether thermal effect are included or not). This 
finding does not require any further analysis of the linearly polarized applied 
field.

However, the study of energy losses in case of the linearly polarized applied 
field for anisotropic nanoparticles enables us to consider whether the large 
anisotropy can be used to explain the experimental results of \cite{Ahsen2010}. 
Let us note the formation of chains by nanoparticles is a known feature of 
ferrofluids and the chains of particles represents a very large anisotropy \cite{He}. 
Therefore, in this section we discuss the solution of the LLG equations \eq{LLG} 
obtained for a single-domain magnetic nanoparticle under linearly polarized, 
i.e. alternating applied (external) magnetic field in the limit of large anisotropy. 
The applied field is assumed to oscillate along the x-axis with an angular 
frequency $\omega$
\begin{eqnarray}
\label{osc_field}
{\bf H_{\mr{ext}}} &=& 
\frac{\omega_L}{\gamma'} \, \, (\cos(\omega t), 0, 0),
\end{eqnarray}
where $\omega_L = \gamma' \vert{\bf H}\vert$ is the Larmor frequency.
Similarly to the circularly polarized case, here, we consider particles with 
uniaxial anisotropy. The easy axis of the magnetization is chosen to be the 
x-axis (similar results can be obtained if it is chosen to be perpendicular to 
the x-axis)
\begin{eqnarray}
\label{aniso_osc_field1}
{\bf H_{\mr{aniso}}} &=& 
\frac{\omega_L}{\gamma'} \, \, (\lambda_{\mr{eff}} M_x, 0, 0),
\end{eqnarray}
where $M_x$ is the x-component of the magnetization vector and the
parameter $\lambda_{\mr{eff}}$ describes the strength of the anisotropy. Note 
that in case of alternating applied field, it is convenient to study the original 
LLG equation \eq{LLG} instead of the rotated one \eq{rot_LLG_descartes}. 
The LLG equation for the Cartesian coordinates of the magnetization reads as 
\begin{align}
\label{osc_LLG}
&\frac{\mr{d} M_x}{\mr{d}t} = 
\alpha_N (\lambda_{\mr{eff}}  M_x + \cos(\omega t)) (1-M_x^2),
\nonumber \\
&\frac{\mr{d} M_y}{\mr{d}t} = -(\omega_L M_z + \alpha_N M_x M_y)
(\cos(\omega t) + \lambda_{\mr{eff}}  M_x), 
\nonumber \\
&\frac{\mr{d} M_z}{\mr{d}t} = (\omega_L M_y +\alpha_N M_x M_z)
(\cos(\omega t) + \lambda_{\mr{eff}}  M_x),
\end{align}
which has in general no analytic solution. In the limit of extremely large anisotropy, 
$\cos(\omega t) \ll \lambda_{\mr{eff}}  M_x$, however, the time-dependence of 
$M_x$ can be determined as
\begin{align}
\label{sol_osc_LLG}
M_{x}(t) = \pm 
\frac{\exp(\alpha_N \lambda_{\mr{eff}} t)}{\sqrt{\exp(2\alpha_N \lambda_{\mr{eff}} t) - C}},
\end{align}
with $C = 1-1/M_{x0}^2$. If $t\to\infty$ the solution \eq{sol_osc_LLG} 
tends to $\pm 1$ and the energy loss per cycle vanishes. 

We conclude that in case of a very large anisotropy the heat power of a 
magnetic nanoparticle driven by a linearly polarized applied field vanishes. 
The same was observed in case of the circularly polarized external field. 
Thus, it is a natural requirement to obtain a comparable heat power given 
by the linearly and the circularly polarized applied fields if the anisotropy is 
large enough which can explain the experimental results of \cite{Ahsen2010}. 
Indeed, if the ferrofluid was not prepared appropriately, the nanoparticles can 
form chains and consequently, the anisotropy could become large and
the energy loss tends to zero rapidly.

\section{Thermal effects}
\label{sec_thermal}

In this work we studied the influence of anisotropy on the energy losses in 
the framework of the deterministic LLG equation in the absence of thermal 
fluctuations  for rotating applied field. In this subsection, we discuss briefly how 
thermal effects (see e.g. \cite{stochastic_llg_lin,stochastic_llg,JaHaCh2012})
can possibly modify the results obtained by considering purely dynamical 
effects based on the LLG equation. Let us follow Refs. \cite{Denisov2006,
Denisov2006prl,Denisov_thermal}. In case of a rotating field, the steady state 
solutions of the dynamical problem are independent of the initial conditions of 
the individual particles. Therefore, the average magnetization can be easily 
determined by these (stable) steady states. 

In case of two steady state magnetizations (up and down states), 
due to thermal effects, a nonzero probability of a transition from one stable 
state to another appears. This type of relaxation mechanism is missing in 
the present work due to the lack of thermal effects. We showed that large 
anisotropy is needed in order to have more then one stable state. It was 
also shown that if the anisotropy is not large enough only a single fixed 
point appear in the phase portrait, thus one has to consider only a single 
stable state. In this case the modification caused by thermal effects is less 
important. Thus, thermal effects can only modify the determination of energy 
losses (in case of a rotating applied field, for low frequencies relevant to 
hyperthermia) if the anisotropy is large enough but not too large (otherwise 
the barrier between one state to an other becomes too large). 

Furthermore, let us compare \cite{Raikher2011} and \cite{Chatel2009} in 
order to consider the role of thermal effects in case of isotropic samples. In 
these articles energy losses were investigated both for alternating and for 
rotating applied field. Thermal effects were included in \cite{Raikher2011} 
while these are absent in \cite{Chatel2009}. For example, the dashed lines 
on Fig. 6 of  \cite{Raikher2011} correspond to the limit $T\to 0$ and agree to 
the findings plotted in Fig. 2a of \cite{Chatel2009} qualitatively. Let us pay the
attention of the reader to the logarithmic scale used in Fig. 2a of \cite{Chatel2009}. 
The modification caused by thermal effects (solid lines in Fig. 6) are less important 
for the rotating case but very significant for the alternating case in the limit of low 
frequencies. Nevertheless, for low frequencies, the linearly polarized field 
was found to produce more heat power both for $T=0$ and for $T \neq 0$. 

According to Refs. \cite{Denisov2006,Denisov2006prl,Denisov_thermal} for 
small and for very large anisotropy the thermal effects are less important. It was 
shown in Ref. \cite{Raikher2011} that for isotropic samples even for $T \neq 0$ 
the energy absorption per cycle for a nanoparticle was larger in case of a linearly 
polarized applied field. Thus, if anisotropy (at least if it is small or very large) 
does not increase the energy losses obtained in case of a rotating external field 
(which is one of the finding of the present work), a better heating efficiency can be 
achieved for isotropic nanoparticles using linearly polarized field. This indicates 
that possible modification caused by thermal effects can only be expected in case 
of moderate anisotropy. Therefore, the study of the present work is relevant for 
applications in hyperthermia at least for small and very large anisotropy.

\section{Summary}
\label{sec_sum}
The nonlinear dynamics of magnetization and the loss energy of a single magnetic 
nanoparticle with uniaxial anisotropy have been considered under circularly 
polarized applied field in the absence of thermal fluctuations. The easy axis of 
magnetization has been chosen to be perpendicular to the rotating applied field. 
We solved the deterministic Landau-Lifshitz-Gilbert equation in order to determine 
the loss energy per cycle in case of the rotating applied field and the findings were 
compared to that of obtained for the isotropic case in \cite{Chatel2009}. Comparison 
between the linearly and circularly polarized applied field has also been performed 
and the results were analyzed in terms of the experimental data \cite{Ahsen2010}. 

Our goal was twofold: (i) to study whether the anisotropy can be used to achieve a 
better heating efficiency in case of rotating external field, (ii) to use the anisotropy 
to resolve the discrepancy between theory \cite{Chatel2009} and experiment
\cite{Ahsen2010}. We showed that for circularly polarized field, the loss energy per 
cycle is decreased by the anisotropy compared to the isotropic case. Thus in the 
low-frequency limit, more heat power can be achieved by alternating applied field 
for isotropic nanoparticles, at least the rotating applied field produces lower energy 
absorption for small and very large anisotropy. It was also shown that in the limit of 
extremely large anisotropy, experimental results of \cite{Ahsen2010} can be 
explained. The possible role of thermal fluctuations discussed here indicates the 
necessity of the extension of the present study for the case of moderate anisotropy 
when thermal effects are taken into account appropriately.

\section*{Acknowledgement}
This research was supported by the T\'AMOP 4.2.1./B-09/1/KONV-2010-0007 project.
Fruitful discussions with P.F. de Ch\^atel, J. Hakl, Zs. J\'anosfalvi, S. Nagy and K. Vad 
are warmly acknowledged.

\end{document}